  \documentclass[preprint]{aastex}

\received{2001 March 20}
\accepted{2001 April 17}

\slugcomment{To appear in the {\em Astrophysical Journal Letters.}}

\shorttitle{Inner Light Year of Seyfert Nucleus in NGC\,4395}
\shortauthors{Wrobel, Fassnacht, \& Ho} 

\begin{document}

\title{The Inner Light Year of the Nearest Seyfert~1 Nucleus in NGC\,4395}
\author{J.~M. Wrobel and C.~D. Fassnacht\altaffilmark{1}}
\affil{National Radio Astronomy Observatory,
       P.O. Box O, Socorro, New Mexico 87801}
\and
\author{L.~C. Ho}
\affil{The Observatories of the Carnegie Institution of Washington,\\
       813 Santa Barbara Street, Pasadena, CA 91101}
\email{jwrobel@nrao.edu, cfassnacht@nrao.edu, lho@ociw.edu}

\altaffiltext{1}{Current address: Space Telescope Science 
Institute, 3700 San Martin Drive, Baltimore, MD 21218}

\begin{abstract}
The NRAO VLBA was used at 1.4~GHz to image the inner 550~mas
(23~lt-years) of the nearest known Seyfert~1 nucleus, in the Sm galaxy
NGC\,4395.  One continuum source was detected, with flux density
$530\pm130~\mu$Jy, diameter $d < 11$~mas (0.46~lt-year), and
brightness temperature $T_{\rm b} > 2.0\times10^6$~K.  The spectral
power $P\/$ of the VLBA source is intermediate between those of
Sagittarius\,A and Sagittarius\,A$^*$ in the Galactic Center.  For the
VLBA source in NGC\,4395, the constraints on $T_b$, $d\/$, and $P\/$
are consistent with an origin from a black hole but exclude an origin
from a compact starburst or a supernova remnant like Cassiopeia\,A.
Moreover, the spectral powers of NGC\,4395 at 1.4 and 4.9~GHz appear
to be too low and too constant to allow analogy with SN\,1988Z, a
suggested prototype for models of compact supernova remnants.  The
variable and warm X-ray absorber in NGC\,4395 has a free-free optical
depth much larger than unity at 1.4~GHz and, therefore, cannot fully
cover the VLBA source.
\end{abstract}

\keywords{galaxies: active - 
          galaxies: individual (NGC\,4395, UGC\,07524) - 
          galaxies: nuclei - 
          galaxies: Seyfert - 
          radio continuum: galaxies - 
          X-rays: galaxies}

\section{MOTIVATION}

The Seyfert~1 nucleus of NGC\,4395 holds three unique distinctions.
First, in the optical regime it is the least luminous Seyfert nucleus
known, with an absolute blue magnitude of only $M_B=-9.8$~mag.  On
energetics grounds it is therefore essential to explore both black
hole and stellar origins for the Seyfert activity
\citep{fil89,fil93,lir99}.  Second, it is the only Seyfert nucleus
known to be hosted by an Sm galaxy.  Proving that the Seyfert traits
of NGC\,4395 are inconsistent with a stellar origin, but consistent
with a black hole origin, would thus bolster claims of black hole
ubiquity in the local universe \citep{mag98}.  Finally, NGC\,4395 lies
at a distance $D=2.6$~Mpc \citep{row85}, making it the nearest known
Seyfert~1 nucleus.  The corresponding scale is 10~mas $=$
0.42~lt-year, so the inner light year can, in theory, be probed
directly with the NRAO Very Long Baseline Array (VLBA).  Direct
imaging of radio emission on light-year scales can be a powerful
discriminant between black hole and stellar origins: a black hole
could launch jets \citep{fal00}, while stellar processes could result
in a compact starburst \citep{con91}, a supernova remnant resembling
Galactic ones \citep{gre84}, or a compact supernova remnant
\citep{ter95}.

Recent observations of NGC\,4395 with the NRAO Very Large Array (VLA)
showed an unresolved source at a frequency $\nu=1.4$~GHz, with a flux
density $S=1680\pm94~\mu$Jy; a diameter $d < 550$~mas (23~lt-years);
and a spectral index $\alpha = -0.60\pm0.08$ ($S \propto
\nu^{\alpha}$) between 1.4 and 4.9~GHz, indicating synchrotron
emission from a nonthermal plasma \citep{ho01a}.  A VLA source of this
strength is too weak to be imaged with the VLBA if just traditional
self-calibration techniques are applied.  However, the prospects for
successfully imaging NGC\,4395 with the VLBA are good if
phase-referencing techniques are employed \citep{wro00b}.  Section~2
of this Letter reports the detection of one VLBA source at 1.4~GHz,
using phase-referenced observations of a 550-mas region in NGC\,4395.
Section~3 examines the implications of this VLBA detection, plus
published radio photometry, for black hole and stellar models for the
origin of the radio emission.  These data are shown to be consistent
with an origin from a black hole but inconsistent with an origin from
a compact starburst, a supernova remnant like Cassiopeia\,A, or a
compact supernova remnant resembling SN\,1988Z.

\section{OBSERVATIONS, CALIBRATION, AND IMAGING}

The VLBA \citep{nap94} was used to observe NGC\,4395 and calibrators
on 2000 April 11 UT.  Data were acquired in dual circular
polarizations with 4-level sampling and at a center frequency
1.43840~GHz with bandwidth 32~MHz.  Phase-referenced observations were
made in the nodding style.  A 3-minute observation of NGC\,4395 was
preceded and followed by a 2-minute observation of the phase, rate,
and delay calibrator J1220+3431 \citep{wil98} about $1.5\arcdeg$ from
NGC\,4395.  Sources J1215+3448 and J1310+3220 were also observed,
respectively, to check the astrometric accuracy and to align the
phases of the independent baseband channels \citep{ma98}.  Observation
and correlation assumed a coordinate equinox of 2000.  The {\em a
priori\/} position adopted for NGC\,4395 was very close to that in
\citet{ho01a}.

Data editing and calibration were done using the 1999 December 31
release of the NRAO AIPS software and following the strategies
outlined by \citet{ulv00}.  After data deletion based on {\em a
priori\/} flags, data were deleted on all baselines involving any
antenna observing below an elevation of $20\arcdeg$.  Such
elevation-based editing minimized differential ionospheric conditions
between the lines of sight to J1220+3431 and NGC\,4395.  Despite this
step, it was impossible to calibrate the phases on baselines to the
VLBA antenna on Manua Kea in Hawaii, and those baselines were also
deleted.  This resulted in an observed baseline range of 240--5800~km
and, for NGC\,4395, a total of 129~baseline-hours of integration.
VLBA system temperatures and gains were used to set the amplitude
scale to an accuracy of about 5\%, after first correcting for sampler
errors.  No self-calibrations were performed on NGC\,4395.

The AIPS task IMAGR was used to form and deconvolve an image of the
Stokes $I\/$ emission from NGC\,4395, with the visibility data being
naturally weighted to optimize image sensitivity.  This image, given
in Figure~1, was restored with an elliptical-Gaussian beam with FWHM
dimensions of 13.9~mas (0.59~lt-year) by 10.7~mas (0.45~lt-year) and
elongation orientation of $-20.3\arcdeg$, and has an rms noise value
of 51~microjanskys ($\mu$Jy) per beam area.  The left panel of
Figure~1 shows that only one VLBA source was detected above 4.5 times
the rms noise, within a region centered on the unresolved VLA
detection and spanning 550~mas (23~lt-years), which matches the upper
limit to the diameter of the VLA detection at 1.4~GHz \citep{ho01a}.
A quadratic fit to the peak of the VLBA detection yielded a position
of $\alpha(J2000) = 12^{h} 25^{m} 48^{s}.874$ and $\delta(J2000) =
33^{\circ} 32' 48''.69$.  This position carries an absolute
2-dimensional error of 55~mas, set by the position error of J1220+3431
and verified with a phase-referenced image of J1215+3448.  Given the
signal-to-noise ratio of the VLBA detection of NGC\,4395, its position
can, in principal, be determined with an error of 1-2~mas
\citep{bal75} and efforts are underway to reach that limiting
accuracy.

\placefigure{fig1}

The right panel of Figure~1 displays the inner 76~mas (3.2~lt-years)
centered on the position of the VLBA detection.  The VLBA source seems
to be slightly resolved but this apparent resolution is likely to be
artifical, due to the weakness of the source and/or residual errors in
the phase calibration.  As a gauge of the importance of the latter
effect, before self-calibration an image of the strong source
J1215+3448 had a peak intensity only 88\% of the peak after phase
self-calibration.  For the VLBA detection of NGC\,4395, (1) image
integration over $N=5.8$ beam areas gives a total flux density of
$530\pm130~\mu$Jy, where the error is the quadratic sum of a 5\% scale
error and $\sqrt{N}$ times the rms noise; and (2) a conservative
estimate to the diameter is $d < 11$~mas (0.46~lt-year).  Traits (1)
and (2) imply that the VLBA detection has a brightness temperature
$T_{\rm b} > 2.0\times10^6$~K.

\section{IMPLICATIONS}

Figure~2 shows that the VLA detections of NGC\,4395 \citep{ho01a}
match the spectral powers and spectral index of Sagittarius\,A in the
Galactic Center \citep{ped89}.  The VLA size limit for NGC\,4395,
conveyed by the dimension of the left panel of Figure~1, is also
comparable to the diameter of 26~lt-years for Sagittarius\,A
\citep{ped89}.  But the VLBA detection of NGC\,4395 at 1.4~GHz is less
powerful (Figure~2) and more compact (Figure~1, left panel) than
Sagittarius\,A.  This hints that the VLBA source in NGC\,4395 could be
an extragalactic analog to Sagittarius\,A$^*$ \citep{fal98}, thought
to mark the massive black hole at the dynamical center of the Galaxy
\citep{rei99}.  An accurate position is not available for the
dynamical center of NGC\,4395 \citep{swa99}.  The position of the VLBA
detection does agree with positions for the optical continuum
\citep{cot99}, X-ray continuum \citep{ho01b}, and nuclear H$\alpha$
emission \citep{van98}, but these agreements carry combined errors of
order 1000~mas (42~lt-years).  However, a stringent mass limit is
available for a putative black hole in NGC\,4395: \citet{fil01} report
detection of the \ion{Ca}{2} infrared triplet lines in absorption from
echelle spectra taken with the Keck~I telescope, yielding estimates
for the strength of the stellar contribution to the nuclear light
($M_B=-7.3$~mag) and the central LOS velocity dispersion ($\sigma \sim
30$~km~s$^{-1}$), which, in combination with the star cluster size
from {\em HST\/} images, limit any black hole mass to $M_{\rm BH}
\lesssim 80,000~M_{\sun}$.

\placefigure{fig2}

An accreting black hole could launch jets.  While \citet{fal00} use a
spectral analysis to build a case for a jet origin for
Sagittarius\,A$^*$, morphological evidence for an outflow would be far
more compelling.  In the case of NGC\,4395, the VLBA source is
unresolved at a linear resolution of 0.46~lt-year or 170~lt-days, and
only 0.32$\pm$0.26 times as strong as the VLA source spanning
23~lt-years or less.  A few nearby galaxies have had their radio
nuclei probed on similar scales, or even finer scales down to a
resolution of 10~lt-days (eg, NGC\,3031, Bietenholz, Bartel, \& Rupen
2000).  While jets or jet-like structures are invariably observed,
those galaxies exhibit spectral powers at 1.5~GHz that are an order of
magnitude, or more, above the power of the source in NGC\,4395 at
comparable VLA resolutions \citep{wro01}.  Still, the VLA and VLBA
sources in NGC\,4258 at 1.5~GHz \citep{cec00} are only about a factor
of ten more powerful than their counterparts in NGC\,4395.  At a
linear resolution of 0.94~lt-year, the VLBA source in NGC\,4258 is
resolved into two jets elongated over 5.9~lt-years.  Similar jets, if
present in NGC\,4395, could account for some of the VLA flux density
missing from Figure~1: a deeper VLBA image of NGC\,4395 is required to
search for such jets.  The present VLBA detection could then
correspond to the brightest region in the jets, an hypothesis that
could be tested with VLBA imaging at higher linear resolution.

On energetics grounds it is reasonable also to explore a stellar
origin for the Seyfert activity in NGC\,4395 \citep{fil93,lir99}.  In
the radio regime, stellar processes could result in a compact
starburst, a supernova remnant resembling Galactic ones, or a compact
supernova remnant.

For a compact starburst, \citet{con91} use an empirical scaling
relation between thermal and nonthermal radio continuum, based on more
extended star-forming galaxies with thermal electron temperature
$T_{\rm e} = 10^4$~K, to derive a limiting brightness temperature of
$T_{\rm b} \lesssim 10^5$~K at 1.4~GHz.  The VLBA source in NGC\,4395,
with brightness temperature $T_{\rm b} > 2.0\times10^6$~K, clearly
exceeds this limit, excluding a compact starburst origin.

The Galactic supernova remnant Cassiopeia\,A, of age $t \lesssim
400$~years, has spectral powers and a spectral index
\citep{baa77,gre84,fil93} very similar to those derived from the VLA
detections of NGC\,4395 \citep{ho01a}.  These photometric similarities
are displayed in Figure~2.  The upper limit $d = 23$~lt-years to the
diameter of the VLA source in NGC\,4395 is also consistent with the
diameter $d = 13$~lt-years for Cassiopeia\,A \citep{gre84}.  But the
VLBA source in NGC\,4395 is about a third as powerful (Figure~2) and
considerably more compact ($d < 0.46$~lt-year) than Cassiopeia\,A.
This VLBA size constraint strongly excludes further analogy with
Cassiopeia\,A.

Could the VLBA source in NGC\,4395 be a compact supernova remnant
(cSNR) whose evolution is governed by a supernova expanding into
dense circumstellar material \citep{ter95}?  Such an interpretation
encounters some successes and some difficulties \citep{fil93,lir99}.
In the optical regime, published images of NGC\,4395 plus the POSS
revealed the presence of a starlike nucleus, visible at similar
brightness levels since 1956 May 8 UT and implying an age $t \gtrsim
36$~years in 1992.  \citet{lir99} find that, while this slow blue
photometric evolution could just be a signpost of a cSNR of age $t
\sim 300$~years, the observed line properties roughly match those
expected for a cSNR of age $t \sim 34$~years.  The large-amplitude
variability observed on time scales of days to months, in both the
optical and X-ray regimes, remains a problem for the cSNR model.

In the radio regime, the record of photometric evolution of the VLA
source, while sparse, supports approximate constancy over 1--2 decades
at 1.4 and 4.9~GHz, since 1982 \citep{mor99,ho01a}.  Also,
\citet{hee64} used the NRAO 91-m telescope in 1963 to set an upper
limit of 0.2~Jy for the peak flux density of NGC\,4395 at 1.4~GHz.  Is
this record consistent with a cSNR origin?  Models for cSNRs do not,
as yet, predict radio light curves, although \citet{ter95} do remark
on the relevance of SN\,1988Z, a radio supernova, to their models.
Figure~3 shows the model light curves for SN\,1988Z at 1.4 and
4.9~GHz, based on three years of monitoring in the radio regime
\citep{van93} but extrapolated to an age of 50~years.  The powers of
the VLA detections of NGC\,4395, measured on 1982 Feb 8 UT and 1990
Mar 3 UT by \citet{mor99} with matched resolutions, and on 1999 Aug 29
and Oct 31 UT by \citet{ho01a} also with matched resolutions, are
plotted at their minimum possible ages assuming a reference date of
1956 May 8 UT.  (No point appears for the VLBA detection due to
concern about the effects of source resolution between VLA and VLBA
scales.)  The power of the 91-m upper limit, obtained during 1963
between Feb 1 UT and Dec 31 UT, is also plotted at the minimum
possible age of about 7.2~years relative to the same reference date.
Figure~3 shows that NGC\,4395 is observed to be less powerful than
SN\,1988Z, by factors ranging from at least 60 at an age of about
7~years to about 100--700 at an age of 43~years.  This latter
discrepancy could reflect the failure of the SN\,1988Z model after a
few decades but model problems are unlikely through the first decade
\citep{hym95}.  The spectral powers of NGC\,4395 at 1.4 and 4.9~GHz
thus appear to be too low since 1963 and too steady since 1982 to
allow analogy with SN\,1988Z.  The former trait is more constraining
than the latter, as the degree of variability could be artificially
suppressed if only a small fraction of the VLA emission arises from a
cSNR.

\placefigure{fig3}

In summary, for the VLBA source in NGC\,4395, the constraints on
$T_b$, $d\/$, and $P\/$ are consistent with an origin from a black
hole but exclude an origin from a compact starburst or a supernova
remnant like Cassiopeia\,A.  Also, the spectral powers of NGC\,4395 at
1.4 and 4.9~GHz seem both too low and too stable for analogy with
SN\,1988Z, a suggested prototype for models of compact supernova
remnants.  Future studies of the VLBA source in NGC\,4395 should focus
on measuring its spectral index, reducing the upper limit to its
diameter or seeking evidence for jet-like structures, assessing its
astrometric stability (cf.\ Wrobel 2000), and imaging the missing VLA
flux density.

Like other Seyfert galaxies, NGC\,4395 exhibits copious evidence for
thermal nuclear plasma.  Estimates of the associated free-free optical
depths $\tau_{\rm ff}$ at 1.4~GHz can constrain the relative
geometries of the thermal and nonthermal plasmas.  Equation~1 of
\citet{ulv99} reduces to $\tau_{\rm ff} = 0.038~T_{\rm e}^{-1.35}~E$
at 1.4~GHz for a thermal plamsa with electron temperature $T_{\rm e}$
in units of K and emission measure $E\/$ in units of cm$^{-6}$~pc.
While $T_{\rm e}$ is either readily observed or plausibly estimated,
neither condition generally applies to $E$.  However, a notable
exception is the thermal plasma responsible for the variable warm
X-ray absorber in NGC\,4395 \citep{iwa00}.  For this plasma, $T_{\rm
e} = 10^6$~K is plausibly adopted, the column density is measured, and
the variability time scale sets a lower limit to the density.  The
latter two quantities imply $E \gtrsim 2 \times 10^{13}$~cm$^{-6}$~pc,
which, in combination with $T_{\rm e} = 10^6$~K, leads to $\tau_{\rm
ff} \gtrsim 6000$.  The variable warm X-ray absorber in NGC\,4395 has
a free-free optical depth much larger than unity at 1.4~GHz and,
therefore, cannot fully cover the VLBA source.  Arguments such as
these should be folded into discussions of unifying structures for the
inner regions of quasars (eg, Elvis 2000).

\acknowledgments The authors thank Dr.\ J.\ Ulvestad for discussions.
This research has made use of the NASA/IPAC Extragalactic Database
(NED) which is operated by the Jet Propulsion Laboratory, Caltech,
under contract with the National Aeronautics and Space Administration.
NRAO is a facility of the National Science Foundation operated under
cooperative agreement by Associated Universities, Inc.

\clearpage

\figcaption{VLBA images of Stokes $I\/$ emission from NGC\,4395 at
a frequency of 1.4~GHz.  Hatched ellipse shows the restoring beam area
at FWHM.  Contour levels are $\pm$2, $\pm$4, and $\pm$6 times the
image rms noise level of 51~$\mu$Jy per beam area.  Negative contours
are dashed and positive ones are solid.
{\em Left:} Inner 550~mas (23~lt-years) centered on the position of
the VLA detection at 1.4~GHz by Ho \& Ulvestad (2001).  The image size
matches the upper limit to the diameter of the unresolved VLA
detection.  One-sigma position errors are shown by the cross for the
VLA detection and by the cross with a gap for the VLBA detection.
{\em Right:} Inner 76~mas (3.2~lt-years) centered on the position of
the VLBA detection.
\label{fig1}}

\figcaption{Spectral power as a function of frequency.  One-sigma
error bars are shown for NGC\,4395.  Error bars for the Galactic
sources are smaller and not shown to reduce plot clutter. 
\label{fig2}}

\figcaption{Spectral powers at 1.4~GHz (circles) and 4.9~GHz (squares)
as a function of age.  The model light curves for SN\,1988Z, shown at
monthly intervals, are from three years of monitoring in the radio
regime but extrapolated to an age of 50~years.  The data points are
from measurements of NGC\,4395 during 1963--1999 but referenced to
1956 May 8 UT, the epoch of the first optical detection of the
Seyfert~1 nucleus.  One-sigma error bars are shown for NGC\,4395.
\label{fig3}}

%
\newpage 
  \epsscale{1.0} \plottwo{f1a.eps}{f1b.eps}

  \centerline{Figure~1}

\newpage 
  \epsscale{2.0} \plotone{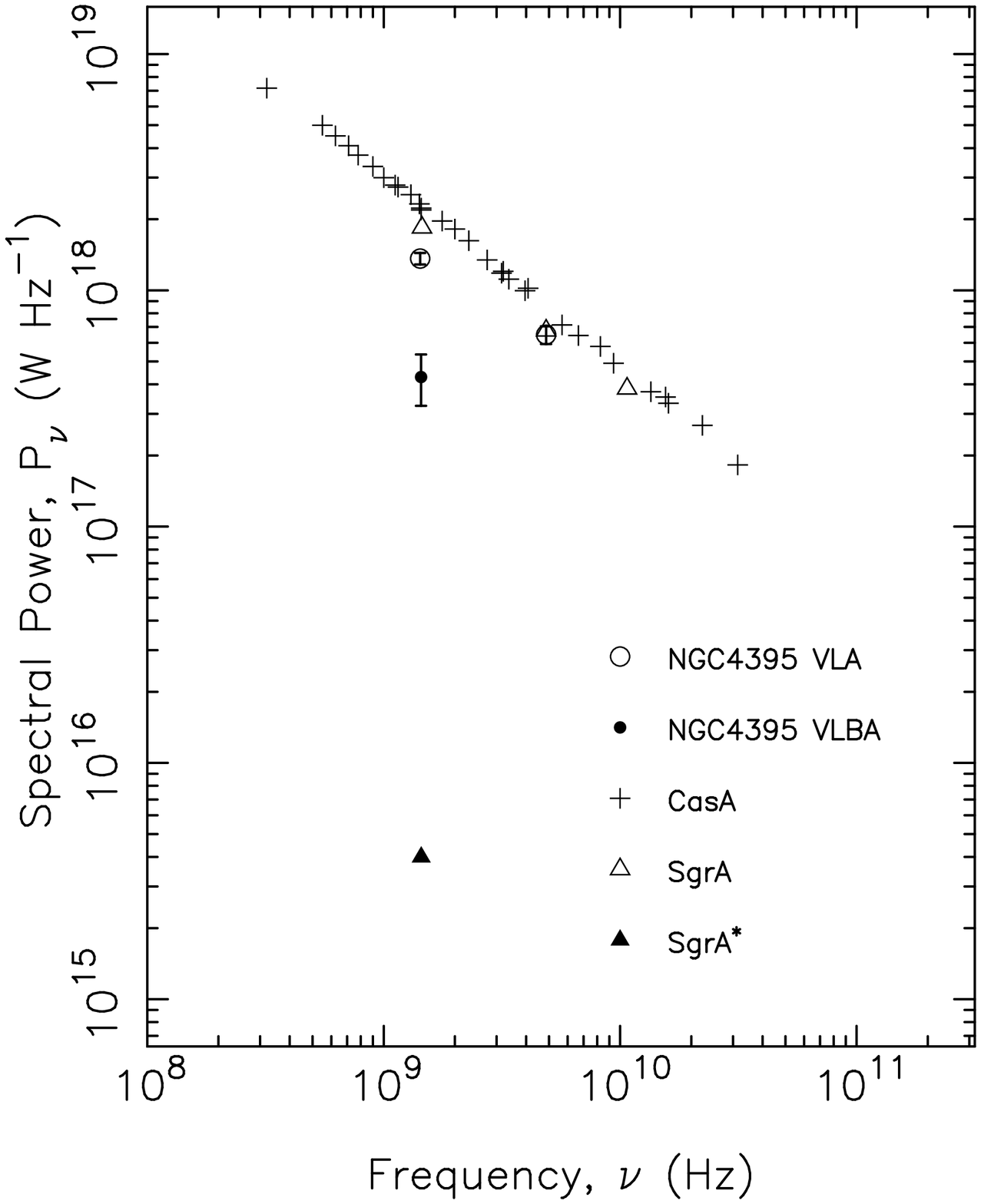}

  \centerline{Figure~2}

\newpage 
  \epsscale{1.7} \plotone{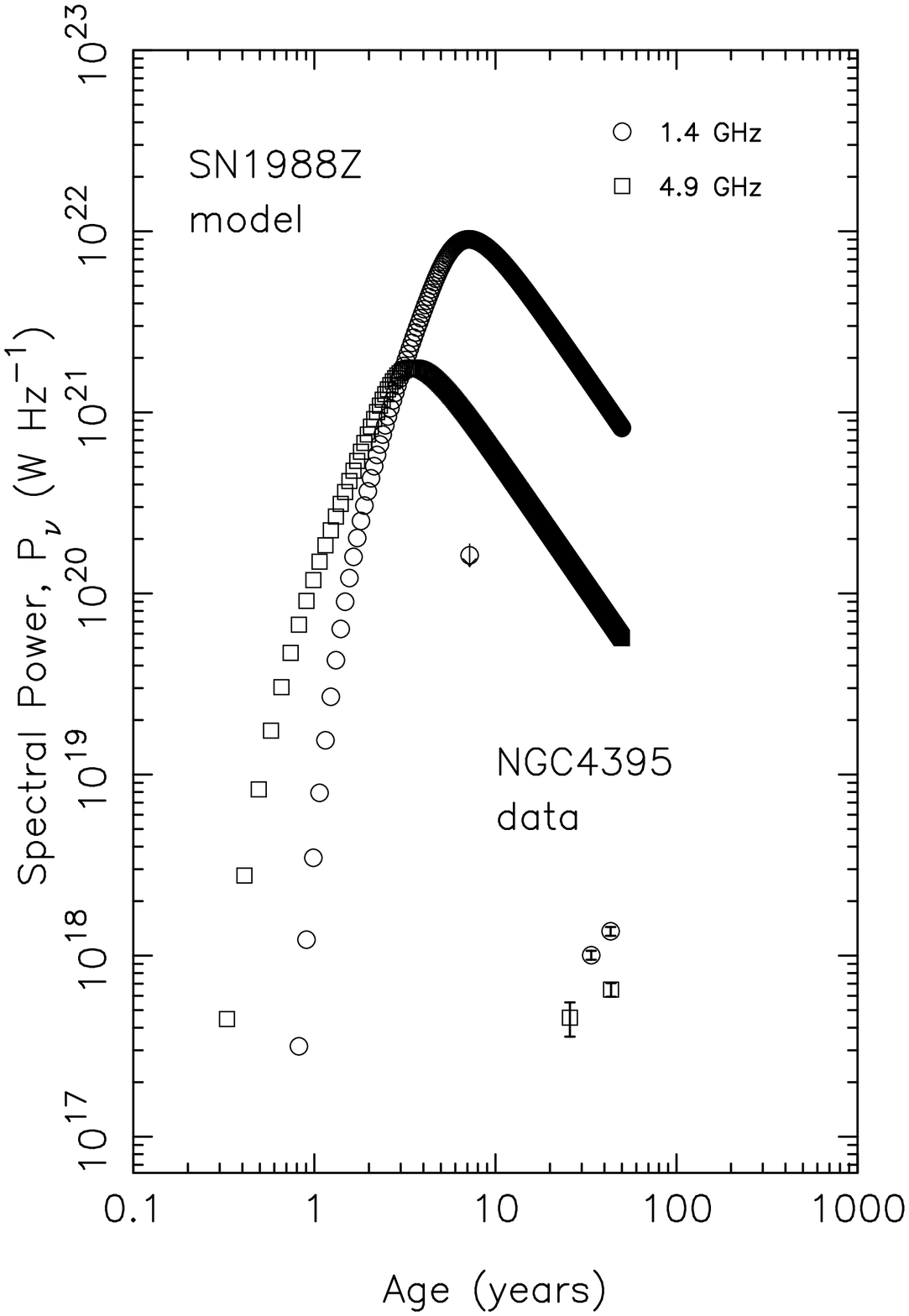}

  \centerline{Figure~3}


\clearpage

\begin{thebibliography}{}
\bibitem[Baars et al.(1977)]{baa77}
   Baars, J.~W.~M., Genzel, R., Pauliny-Toth, I.~I.~K., \& Witzel, A.
   1977, \aap, 61, 99
\bibitem[Ball(1975)]{bal75} 
   Ball, J.~A. 1975, in Methods in Computational Physics, 
   Volume 14, eds.\ B. Alder, S. Fernbach, \& M. Rotenberg 
   (New York: Academic Press), 177
\bibitem[Bietenholz, Bartel, \& Rupen(2000)]{bie00} 
   Bietenholz, M.F., Bartel, N., \& Rupen, M.P. 2000, \apj, 532, 895
\bibitem[Cecil et al.(2000)]{cec00} 
   Cecil, G., et al. 2000, \apj, 536, 675
\bibitem[Condon et al.(1991)]{con91}
   Condon, J.~J., Huang, Z.-P., Yin, Q.~F., \& Thuan, T.~X. 1991, 
   \apj, 378, 65
\bibitem[Cotton, Condon, \& Arbizzani(1999)]{cot99} 
   Cotton, W.~D., Condon, J.~J., \& Arbizzani, E. 1999, \apjs, 125, 
   409
\bibitem[Elvis(2000)]{elv00}
   Elvis, M. 2000, \apj, 545, 63
\bibitem[Falcke et al.(1998)]{fal98} 
   Falcke, H., Goss, W.~M., Matsuo, H., Teuben, P., Zhao, J.-H., \&
   Zylka, R. 1998, \apj, 499, 731
\bibitem[Falcke \& Markoff(2000)]{fal00} 
   Falcke, H., \& Markoff, S. 2000, \aap, 362, 113
\bibitem[Filippenko \& Sargent(1989)]{fil89}
   Filippenko, A.~V., \& Sargent, W.~L.~W. 1989, \apj, 342, L11
\bibitem[Filippenko, Ho, \& Sargent(1993)]{fil93}
   Filippenko, A.~V., Ho, L.~C., \& Sargent, W.~L.~W. 1993, \apj, 410, 
   L75
\bibitem[Filippenko \& Ho(2001)]{fil01}
   Filippenko, A.~V., \& Ho, L.~C. 2001, \apj, submitted
\bibitem[Green(1984)]{gre84}
   Green, D.~A. 1984, \mnras, 209, 449
\bibitem[Heeschen \& Wade(1964)]{hee64} 
   Heeschen, D.~S., \& Wade, C.~M. 1964, \aj, 69, 277
\bibitem[Ho \& Ulvestad(2001)]{ho01a} 
   Ho, L.~C., \& Ulvestad, J.~S. 2001, \apjs, 133, 77
\bibitem[Ho et al.(2001)]{ho01b} 
   Ho, L.~C., et al.\ 2001, \apj, 549, L51
\bibitem[Hyman et al.(1995)]{hym95} 
   Hyman, S.~D., Van Dyk, S.~D., Sramek, R.~A., \& Weiler, K.~W. 1995, 
   \apj, 443, L77
\bibitem[Iwasawa et al.(2000)]{iwa00} 
   Iwasawa, K., Fabian, A.~C., Almaini, O., Lira, P., Lawrence, A.,
   Hayashida, K., \& Inoue, H. 2000, \mnras, 318, 879
\bibitem[Lira et al.(1999)]{lir99} 
   Lira, P., Lawrence, A., O'Brien, P., Johnson, R.~A., Terlevich, R., 
   \&  Bannister, N. 1999, \mnras, 305, 109
\bibitem[Ma et al.(1998)]{ma98} 
   Ma, C., et al.\ 1998, \aj, 116, 516
\bibitem[Magorrian et al.(1998)]{mag98} 
   Magorrian, J., et al.\ 1998, \aj, 115, 2285
\bibitem[Moran et al.(1999)]{mor99} 
   Moran, E.~C., Filippenko, A.~V., Ho, L.~C., Shields, J.~C.,
   Belloni, T., Comastri, A., Snowden, S.~L., \& Sramek, R.~A. 1999, 
   \pasp, 111, 801
\bibitem[Napier et al.(1994)]{nap94} 
   Napier, P.~J., Bagri, D.~S., Clark, B.~G., Rogers, A.~E.~E., 
   Romney, J.~D., Thompson, A.~R., \& Walker, R.~C. 1994, Proc.\ IEEE, 
   82, 658
\bibitem[Pedlar et al.(1989)]{ped89} 
   Pedlar, A., Anatharmaiah, K.~R., Ekers, R.~D., Goss, W.~M., van 
   Gorkom, J.~H., Schwarz, U.~J., \& Zhao, J.-H. 1989, \apj, 342, 769
\bibitem[Reid et al.(1999)]{rei99}
   Reid, M.~J., Readhead, A.~C.~S., Vermeulen, R.~C., \& Treuhaft, 
   R.~N. 1999, \apj, 524, 816
\bibitem[Rowan-Robinson(1985)]{row85} 
   Rowan-Robinson, M. 1985, The Cosmological Distance Ladder (New
   York: Freeman)
\bibitem[Swaters et al.(1999)]{swa99} 
   Swaters, R.~A., Schoenmakers, R.~H.~M., Sancisi, R., \& van
   Albada, T.~S. 1999, \mnras, 304, 330
\bibitem[Terlevich et al.(1995)]{ter95} 
   Terlevich, R., Tenorio-Tagle, G., Rozyczka, M., Franco, J., \& 
   Melnick, J. 1995, \mnras, 272, 198
\bibitem[Ulvestad et al.(1999)]{ulv99} 
   Ulvestad, J.~S., Wrobel, J.~M., Roy, A.~L., Wilson, A.~S., Falcke, 
   H., \& Krichbaum, T.~P. 1999, \apj, 517, L81
\bibitem[Ulvestad(2000)]{ulv00}
   Ulvestad, J.~S. 2000, VLBA Scientific Memorandum 25
\bibitem[Van Dyk et al.(1993)]{van93}
   Van Dyk, S.~D., Weiler, K.~W., Sramek, R.~A., \& Panagia, N. 1993,
   \apj, 419, L69
\bibitem[van Zee et al.(1998)]{van98} 
   van Zee, L., Salzer, J.~J., Haynes, M.~P., O'Donoghue, A.~A., \& 
   Balonek, T.~J. 1998, \aj, 116, 2805
\bibitem[Wilkinson et al.(1998)]{wil98}
   Wilkinson, P.~N., Browne, I.~W.~A., Patnaik, A.~R., Wrobel, J.~M., 
   \& Sorathia, B. 1998, \mnras, 300, 790
\bibitem[Wrobel(2000)]{wro00a}
   Wrobel, J.~M. 2000, \apj, 531, 716
\bibitem[Wrobel et al.(2000)]{wro00b}
   Wrobel, J.~M., Walker, R.~C., Benson, J.~M, \& Beasley, A.~J., 
   2000, VLBA Scientific Memorandum 24
\bibitem[Wrobel, Machalski, \& Condon(2001)]{wro01}
   Wrobel, J.~M., Machalski, J., \& Condon, J.~J. 2001, in preparation
\end{thebibliography}
\end{document}